\documentclass[12pt]{article}
\usepackage{epsf,
amsmath,amscd,amssymb,graphicx}
\begin{document}

\thispagestyle{empty}
%
\begin{flushright}
TIT/HEP--489 \\
{\tt hep-th/0212175} \\
December, 2002 \\
\end{flushright}
\vspace{3mm}
\begin{center}
{\Large
{\bf Wall Solution with Weak Gravity Limit}
\\
\vspace{2mm}{\bf in Five Dimensional Supergravity}
\\
} 
\vspace{15mm}

\normalsize

  {\large \bf 
  Masato~Arai~$^{a}$}
\footnote{\it  e-mail address: 
arai@fzu.cz
},  
 {\large \bf 
Shigeo~Fujita~$^{b}$}
\footnote{\it  e-mail address: 
fujita@th.phys.titech.ac.jp
},
  {\large \bf 
  Masashi~Naganuma~$^{b}$}
\footnote{\it  e-mail address: 
naganuma@th.phys.titech.ac.jp
},  

~and~~  {\large \bf 
Norisuke~Sakai~$^{b}$}
\footnote{\it  e-mail address: 
nsakai@th.phys.titech.ac.jp
} 

\vskip 1.5em

{ \it $^{a}$
  Institute of Physics, AS CR, 
  182 21, Praha 8, Czech Republic \\
  and \\
  $^{b}$Department of Physics, Tokyo Institute of 
  Technology \\
  Tokyo 152-8551, JAPAN 
}

\vspace{25mm}
%

{\bf Abstract}\\[5mm]
{\parbox{13cm}{\hspace{5mm}
In five-dimensional supergravity, 
an exact solution of BPS wall is found 
for a gravitational deformation of 
the massive Eguchi-Hanson nonlinear sigma model. 
The warp factor decreases for both infinities of the extra dimension. 
Thin wall limit gives the Randall-Sundrum model without fine-tuning 
of input parameters. 
We also obtain wall solutions with warp factors which are flat or 
increasing in one side, by varying 
a deformation parameter of 
the potential. 
}}
\end{center}
\vfill
\newpage
\setcounter{page}{1}
\setcounter{footnote}{0}
\renewcommand{\thefootnote}{\arabic{footnote}}

{\large\bf Introduction} 
\vspace{2mm}
\\
One of the most interesting models in the brane-world scenario is 
given by Randall and Sundrum, where 
the localization of four-dimensional 
graviton \cite{RS2} has been obtained 
by a spacetime metric 
containing a warp factor $e^{2U(y)}$ which decreases 
exponentially for both infinities of the extra dimension 
$y \rightarrow \pm \infty$ 
\begin{eqnarray}
  ds^2=g_{\mu \nu}dx^\mu dx^\nu = e^{2U(y)}\eta_{mn}dx^mdx^n + dy^2,
\label{5dmetric}
\end{eqnarray}
where $\mu, \nu = 0,..,4$, $m, n = 0,1,3,4$ and $y\equiv x^2$. 
They had to introduce both a bulk cosmological 
constant and a boundary cosmological constant, which have 
to be fine-tuned each other. 
Its supersymmetric (SUSY) version 
has been worked out \cite{ABN}, 
and has been argued to help understanding the fine-tuning \cite{BC1}. 

Since the Randall-Sundrum model uses the bulk AdS space, 
it has implications for the AdS/CFT 
correspondence \cite{AdS/CFT}, \cite{Behrndt}. 
For that purpose it is also natural to 
introduce scalar fields forming a smooth wall 
(thick brane) in supergravity theories. 
After an extensive studies of BPS walls in four-dimensional 
supergravity coupled with  chiral scalar 
multiplets \cite{CQR}, an 
exact solution of a BPS wall has recently been constructed 
\cite{EMSS}. 
Studies of domain wall solutions in gauged supergravity theories 
in five dimensions 
revealed that hypermultiplets are needed 
\cite{KalloshLinde} to obtain 
warp factors decreasing for both infinities $y \rightarrow \pm \infty$ 
(infra-red (IR) fixed points in AdS/CFT correspondence) 
 which are appropriate for phenomenology.
It has been shown that the target space of hypermultiplets in 
five-dimensional supergravity theory must be quaternionic K\"ahler 
manifolds \cite{BaggerWitten} in contrast 
to the hyper-K\"ahler target space 
for five-dimensional global SUSY without gravity. 
The gravitational deformations 
have been worked out for massless ${\cal N}=2$
nonlinear sigma models in four dimensions \cite{Galicki}, \cite{IvanovValent}. 
However, massive models, namely models with potential terms 
are needed to obtain domain wall solutions. 
Massive hyper-K\"ahler nonlinear sigma models without gravity 
in four dimensions have been constructed 
in harmonic superspace as well as in ${\cal N}=1$ superfield 
formulation \cite{ANNS}, and have yielded the domain wall solution for the 
Eguchi-Hanson manifold \cite{Eguchi} previously obtained 
in the on-shell component formulation \cite{Townsend}. 
Domain walls in massive quaternionic K\"ahler nonlinear sigma models 
in supergravity theories have been studied using 
mostly homogeneous target manifolds. 
Unfortunately, SUSY vacua in 
homogeneous target manifolds are not truly IR critical points, 
but can only be saddle points with some IR directions 
\cite{Alekseevsky}, \cite{BC2}. 
Inhomogeneous manifolds and a wall solution have also been 
constructed \cite{Beh-Dall}, \cite{Lazaroiu}. 
However, 
these manifolds do not allow a limit of weak 
gravitational coupling, contrary to the model with an 
exact solution in four-dimensions \cite{EMSS}. 

The purpose of our paper is to present an exact BPS domain wall 
solution in five-dimensional supergravity coupled with hypermultiplets 
(and vector multiplets). 
We have obtained a two-parameter family of massive 
quaternionic nonlinear sigma models 
which reduces to the Eguchi-Hanson nonlinear sigma model 
for vanishing gravitational coupling. 
One of the parameters is the gravitational coupling $\kappa$, and 
the other is an asymmetry parameter $a$ for gravitational 
deformation of potential terms. 
Having a smooth limit of vanishing gravitational coupling is 
very useful to obtain 
inhomogeneous quaternionic K\"ahler manifolds and also 
to use it for brane-world phenomenology. 
The model has two SUSY vacua as genuine local minima up to 
a critical value of gravitational coupling beyond which 
the SUSY vacua become saddle points. 
The BPS domain wall solution for $|a|<1$ gives a warp factor decreasing 
for both infinities of extra dimension $y\rightarrow \pm\infty$, 
interpolating two IR fixed points. 
 For $|a|=1$, the warp factor decreases in one direction, and is flat 
 in the other, interpolating an IR fixed point and flat space. 
 For $|a|>1$, the warp factor decreases in one direction, and 
 increases in the other, interpolating an IR and a ultra-violet (UV) 
 fixed points. 
If we take a thin wall limit for $a=0$, we obtain 
a bulk cosmological constant 
and boundary cosmological constant satisfying the necessary relation 
in Ref.\cite{RS2} 
from our scalar field configuration automatically. 
The relation between two cosmological 
constants is now realized as a consequence of the solution 
of dynamical equations rather than a fine-tuning between 
input parameters, similarly to the BPS wall solution in the 
four-dimensional supergravity \cite{EMSS}. 
Thus we have obtained 
the Randall-Sundrum model as a thin-wall limit of a soliton 
(domain wall) 
in five-dimensional supergravity. 
The four-dimensional 
graviton should be localized on our wall solution \cite{CsabaCsaki}. 

Our strategy to find a gravitational deformation of nonlinear sigma model 
is to use the 
recently obtained off-shell formulation of five-dimensional supergravity 
(tensor calculus) \cite{Fujita-Ohashi}, \cite{FKO} 
combined with the quotient method via a vector multiplet without 
kinetic term and the massive deformation (central charge extension). 
In the off-shell formulation, we can easily introduce the gravitational 
coupling to the massive hypermultiplets with linear kinetic term 
which is interacting with the vector multiplet without kinetic term. 
By eliminating the vector multiplet 
after coupling to gravity, 
we automatically obtain a gravitationally deformed constraint 
resulting in inhomogeneous quaternionic 
K\"ahler nonlinear sigma model with the necessary potential 
terms. 
We may call the procedure a {\it massive} quaternionic 
K\"ahler quotient method. 
If we apply this method to any global ${\cal N}=2$ SUSY model 
with two (or more) isolated SUSY vacua and wall solutions connecting them, 
we should obtain 
a gravitationally deformed inhomogeneous 
quaternionic manifold and wall solutions at least for small gravitational 
coupling.

\vspace{5mm}

{\large\bf Bosonic action of our model in 5D SUGRA} 
\vspace{2mm}
\\
In global ${\cal N}=2$ SUSY case, 
a BPS wall solution in four dimensions has been 
found in the nonlinear sigma model with the Eguchi-Hanson target 
manifold and a potential originating from a mass term 
\cite{Townsend}, \cite{ANNS}. 
Inspired by this solution, we consider a nonlinear sigma model 
of hypermultiplets in five-dimensional supergravity 
which reduces to the massive 
Eguchi-Hanson nonlinear sigma model in the limit of 
vanishing gravitational coupling. 
 For this purpose we use the off-shell formulation 
of Yang-Mills and hypermultiplet matters 
coupled to supergravity in five dimensions 
\cite{Fujita-Ohashi}, \cite{FKO}\footnote{ 
We adopt the conventions of Ref.\cite{Fujita-Ohashi} except 
the sign of our metric $\eta_{\mu\nu}= diag(-1,+1,+1,+1,+1)$. 
This induces a change of 
Dirac matrices 
and the form of SUSY transformation of fermion.
}.
By using the superconformal 
tensor calculus \cite{Fujita-Ohashi}, 
one can obtain the off-shell 
Poincar\'e supergravity action after fixing the extraneous gauge 
freedoms of dilatation, conformal supersymmetry and special 
conformal-boost symmetry \cite{FKO}. 
We start with the system of 
a Weyl multiplet, three hypermultiplets and two $U(1)$ vector multiplets. 
One of the two 
vector multiplets has no kinetic term and plays the role of a Lagrange 
multiplier for hypermultiplets to obtain a curved target manifold. 
The other vector multiplet serves 
to give mass terms for hypermultiplets. 

After integrating out a part of the auxiliary fields 
by their on-shell conditions in 
the off-shell supergravity action \cite{FKO}, 
we obtain the bosonic part of the action 
for our model 
\begin{eqnarray}
   e^{-1}{\cal L}
   &\!\!\!=&\!\!\!
  -\frac{1}{2\kappa^2}R-\frac{1}{4}
  \left(\partial_\mu W_\nu^0-\partial_\nu W_\mu^0 \right)
  \left(\partial^\mu W^{0\nu}-\partial^{\nu} W^{0\mu} \right)
\nonumber \\
   &\!\!\!{}&\!\!\! 
-\nabla^a{\cal A}_i^{\beta}d_\beta{}^\alpha\nabla_a{\cal A}^i_\alpha
    - \kappa^2[{\cal A}^\beta{}_id_\beta{}^\alpha\nabla_a{\cal A}^j_\alpha]^2 
       \nonumber \\
   &\!\!\!{}&\!\!\! -\left[-{\cal A}_i{}^\gamma d_\gamma{}^\alpha 
           (g_0M^0t_0+ M^1t_1)^2{}_\alpha{}^\beta {\cal A}^i_\beta
          -\frac{\kappa^2}{12}(g_0M^0)^2
           (2{\cal A}_\alpha^{(i}d^\alpha{}_\gamma
                (t_0)^{\gamma \beta}{\cal A}^{j)}_\beta)^2 \right],
\label{SUGRA1}
\end{eqnarray}
\begin{eqnarray}
\nabla_\mu{\cal A}_i^{\alpha}
  &=& \partial_\mu {\cal A}_i^\alpha
      -(g_0W^0_\mu t_0+ W^1_\mu t_1)^\alpha{}_\beta {\cal A}_i^\beta , \\
{\cal A}^i{}_\alpha &\equiv & \epsilon^{ij}{\cal A}_j{}^\beta 
\rho_{\beta \alpha} = - ({\cal A}_i{}^\alpha)^* ,
\end{eqnarray}
where $d_\alpha{}^\beta = diag(1,1,-1,-1,-1,-1)$, 
$\kappa$ is the five-dimensional gravitational coupling, 
${\cal A}_i^\alpha, \; i=1,2, \; \alpha=1,\dots,6$ 
are the scalars in hypermultiplets, and 
$W_\mu^0$ ($W_\mu^1$), 
$M^0$ ($M^1$) and $t_0$ ($t_1$) are vector fields, scalar fields 
and generators of the $U(1)$ vector multiplets 
with (without) a kinetic term. 
The gauge coupling of $W_\mu^0$ is denoted by $g_0$. 
Another gauge coupling $g_1$ is absorbed into a normalization of 
$W_\mu^1$ in order to drop the kinetic term by taking 
$g_1 \rightarrow \infty$. 
Hypermultiplet scalars are subject to two kinds of constraints
\begin{eqnarray}
  {\cal A}^2&=&{\cal A}^\beta_id_\beta{}^\alpha{\cal A}^i_\alpha 
   = -2 \kappa^{-2} , 
\label{1stconst} \\
  {1 \over g_1^2}{\cal Y}_1^{ij}
  &\equiv & 2{\cal A}_\alpha^{(i}d^\alpha{}_\gamma
                (t_1)^{\gamma \beta}{\cal A}^{j)}_\beta = 0 .
\label{2ndconst}
\end{eqnarray}
The constraint (\ref{1stconst}) comes from the gauge fixing 
of dilatation, and 
make target space of hypermultiplets to be a non-compact version of 
quaternionic projective space, $\frac{Sp(2,1)}{Sp(2)\times Sp(1)}$, 
combined with the gauge fixing of 
$SU(2)_R$ symmetry
. 
The constraint (\ref{2ndconst}) is required by the on-shell condition 
of auxiliary fields of the $U(1)$ vector multiplet without kinetic term, 
and corresponds to the constraint for Eguchi-Hanson target space 
 in the limit of $\kappa\to 0$. 

The third line of (\ref{SUGRA1}) is a scalar potential 
consisting of two terms : 
the first term arises from the couplings to scalars in vector multiplets 
and the second term from eliminating the auxiliary fields 
of the $U(1)$ vector multiplet with kinetic term. 
The scalar $M^0$ is fixed as 
$(M^0)^2=\frac{3}{2}\kappa^{-2}$ 
from the requirement of canonical normalizations of the Einstein-Hilbert 
term and the kinetic term of 
the gravi-photon $W^0_\mu$ for Poincar\'e supergravity. 
The scalar $M^1$ without kinetic term is a Lagrange multiplier, 
and is found to be 
\begin{eqnarray}
 M^1=-\frac{{\cal A}_i^\gamma d_\gamma{}^\alpha 
     (t_0t_1)_\alpha{}^\beta{\cal A}^i{}_\beta}
     {{\cal A}_i{}^\gamma d_\gamma{}^\alpha(t_1)^2_\alpha{}^\beta
     {\cal A}^i{}_\beta}g_0M^0 .
\end{eqnarray}

Let us introduce two two-component complex fields $\phi_1$ and $\phi_2$ 
to parametrize ${\cal A}_i{}^\alpha$ by a matrix with $i=1, 2$ as rows 
and $\alpha=1,\dots,6$ as columns 
\begin{eqnarray}
 {\cal A}_i{}^\alpha \equiv \frac{1}{\kappa}\bar{\cal A}^{-1/2}\left(
\begin{array}{cccc}
1 & 0 & \kappa\phi_1 & -\kappa\phi_2^* \\
0 & 1 & \kappa\phi_2 & \kappa\phi_1^* 
\end{array} \right) \label{CFbasis} 
\end{eqnarray}
satisfying the constraint (\ref{1stconst}) by 
taking $\bar{\cal A}=1-\kappa^2(|\phi_1|^2+|\phi_2|^2)$. 
 In this basis, 
we can choose two $U(1)$ generators as 
\begin{eqnarray}
 t_1{}^\alpha{}_\beta = \left(
\begin{array}{cccc}
i\alpha & 0 & 0 & 0 \\
0 & -i\alpha & 0 & 0 \\
0 & 0 & i{\bf 1}_2 & 0 \\
0 & 0 & 0 & -i{\bf 1}_2 
\end{array} \right), \quad 
 t_0{}^\alpha{}_\beta = \left(
\begin{array}{cccc}
ia\alpha & 0 & 0 & 0 \\
0 & -ia\alpha & 0 & 0 \\
0 & 0 & -i\sigma_3 & 0 \\
0 & 0 & 0 & i\sigma_3 
\end{array} \right),
\label{generator}
\end{eqnarray}
where $\alpha$ and $a$ are  real parameters and 
$\sigma_3$ is one of the Pauli matrices.
The parameter $\alpha $ in $t_1$ makes target manifold 
inhomogeneous generally through the constraint (\ref{2ndconst}), 
and a special case of $\alpha=1$ corresponds to a homogeneous 
manifold of $SU(2,1)/U(2)$ \cite{BS}.
Here we define $\alpha \equiv \kappa^2\Lambda^3$,
 where $\Lambda$ is a real parameter of unit mass dimension. 
We will show later that this choice of two $U(1)$ 
generators makes 
the hypermultiplet part of this model 
be Eguchi-Hanson 
sigma model with mass term in the limit of $\kappa\to 0$ 
for fixed $\Lambda$. 
The kinetic terms of scalars in hypermultiplets are rewritten as 
\begin{eqnarray}
\frac{1}{2}e^{-1}{\cal L}_{kin}
    &\!\!\!=&\!\!\! 
    -\bar{\cal A}^{-1}[(\partial^\mu \phi_1^*\partial_\mu \phi_1
            + \partial^\mu \phi_2^*\partial_\mu \phi_2) 
            -(|\phi_1|^2+|\phi_2|^2-\kappa^2\Lambda^6)
             W^1_\mu W^{1\mu}]
         \nonumber \\
    &\!\!\!{}&\!\!\!-\kappa^2\bar{\cal A}^{-2}
          [|\phi_2^*\partial_\mu \phi_2+\phi_1\partial_\mu \phi_1^*|^2
           + |\phi_1\partial_\mu \phi_2^*-\phi_2^*\partial_\mu \phi_1|^2] ,
\label{CFkin}
\end{eqnarray}
\begin{eqnarray}
 W^1_\mu=-\frac{{\cal A}^i_\gamma d^\gamma{}_\alpha 
    \overset{\leftrightarrow}{\partial}_\mu
     ((t_1)^\alpha{}_\beta
     {\cal A}_i{}^\beta)}
     {2{\cal A}^i{}_\gamma d^\gamma{}_\alpha(t_1)^{2\alpha}{}_\beta
      {\cal A}_i{}^\beta} 
    =  \frac{i(\phi_1\overset{\leftrightarrow}{\partial}_\mu 
       \phi_1^*+\phi_2
       \overset{\leftrightarrow}{\partial}_\mu \phi_2^*)}
            {2\left(-\kappa^2\Lambda^6+|\phi_1|^2+|\phi_2|^2\right)} ,
\end{eqnarray}
where 
$\phi_1\overset{\leftrightarrow}{\partial}_\mu \phi_1^*
\equiv \phi_1{\partial}_\mu \phi_1^*-({\partial}_\mu \phi_1) \phi_1^*$. 
The constraint (\ref{2ndconst}) becomes 
\begin{equation}
  |\phi_1|^2-|\phi_2|^2 = \Lambda^3,  \qquad 
  \phi_1^*\phi_2 = \phi_2^*\phi_1 = 0.
\label{CFconst}
\end{equation}
After solving the constraint (see Eq.(\ref{Constsolve})) 
 and rewriting the kinetic terms 
 (\ref{CFkin}) by using independent variables, 
 the target metric is found to be a quaternionic extension of the
 Eguchi-Hanson metric \cite{Galicki},~\cite{IvanovValent}.
Since the metric is Einstein, the Weyl tensor is anti-selfdual and 
 the scalar curvature is negative $R=-24\kappa^2$, 
 it is locally a quaternionic manifold \cite{BaggerWitten} 
 for any values of $\kappa \ne 0$.

Potential terms of hypermultiplets become
\begin{eqnarray}
\frac{1}{2}e^{-1}{\cal L}_{pot}
  &\!\!\!=&\!\!\! 
-(g_0M^0)^2
\bar{\cal A}^{-1}
       \left[(-a^2\kappa^2\Lambda^6+|\phi_1|^2+|\phi_2|^2)
       - \frac{[-a\kappa^2\Lambda^6-(\phi_1\sigma_3\phi_1^*
           +\phi_2^*\sigma_3\phi_2)]^2}
              {-\kappa^2\Lambda^6+|\phi_1|^2+|\phi_2|^2}\right] \nonumber \\
  &\!\!\!{}&\!\!\!
+\frac{\kappa^2}{3}(g_0M^0)^2\bar{\cal A}^{-2}
       \left[|\phi_1^*\sigma_3\phi_2+\phi_2\sigma_3\phi_1^*|^2 + 
        |a\Lambda^3 + (\phi_1\sigma_3\phi_1^*-\phi_2^*\sigma_3\phi_2)|^2
       \right] .
\label{CFpot}
\end{eqnarray}
We have now obtained a two-parameter family of gravitational 
deformations of the Eguchi-Hanson metric by means of the gravitational 
coupling $\kappa$ and another deformation parameter $a$ 
specifying the gravitational deformation of potential terms. 
This comes about by an asymmetry of $W^0_\mu t_0$ gauging for 
the central extension (giving mass terms) 
relative to the $W_\mu^1 t_1$ gauging for the constraint 
(producing curved target space of the nonlinear sigma model). 

\begin{figure}[t]
\begin{center}
\leavevmode
\begin{eqnarray*}
\begin{array}{cc}
  \epsfxsize=6cm
  \epsfysize=5cm
\epsfbox{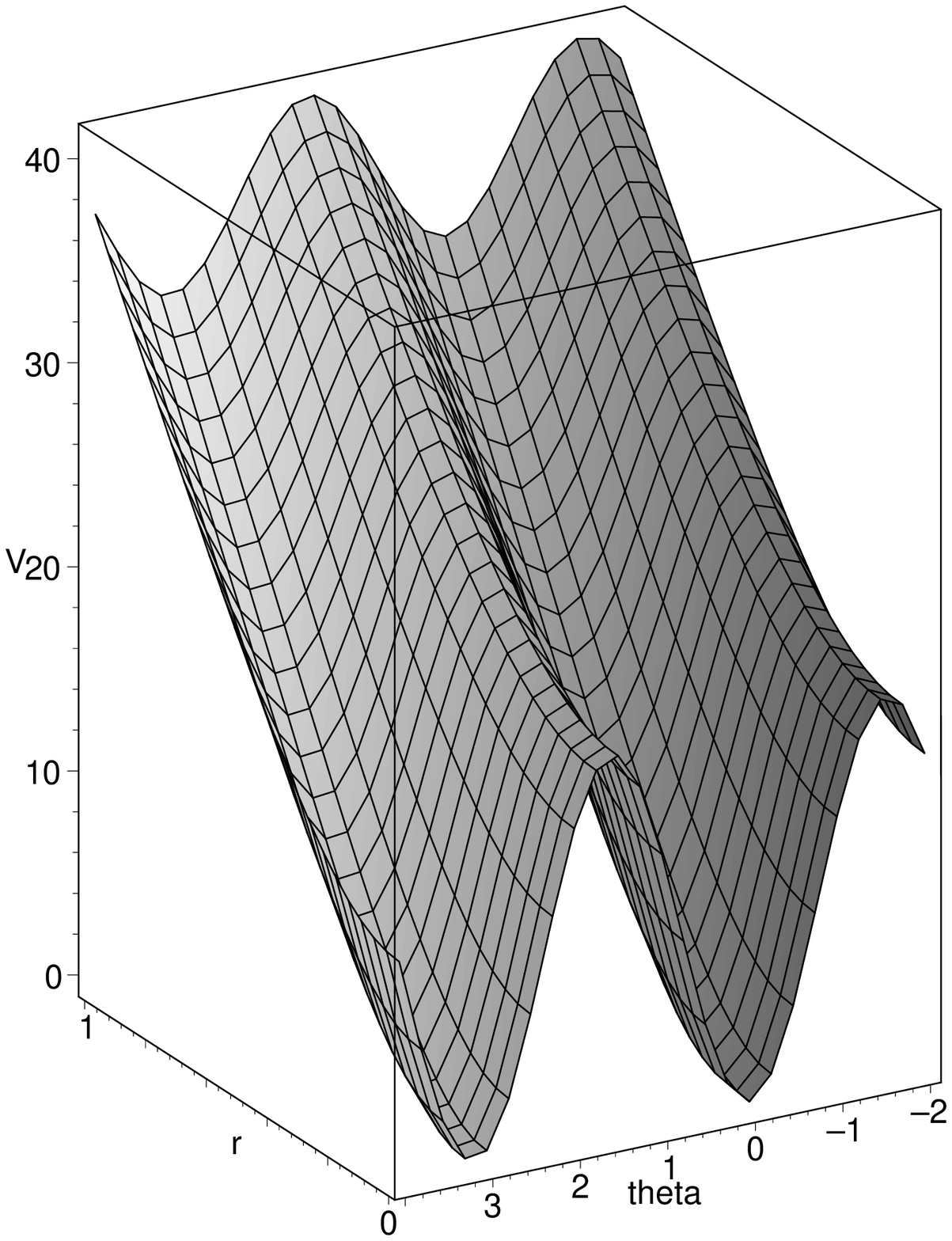} & 
  \epsfxsize=6cm
  \epsfysize=5cm
\hspace*{1cm}
\epsfbox{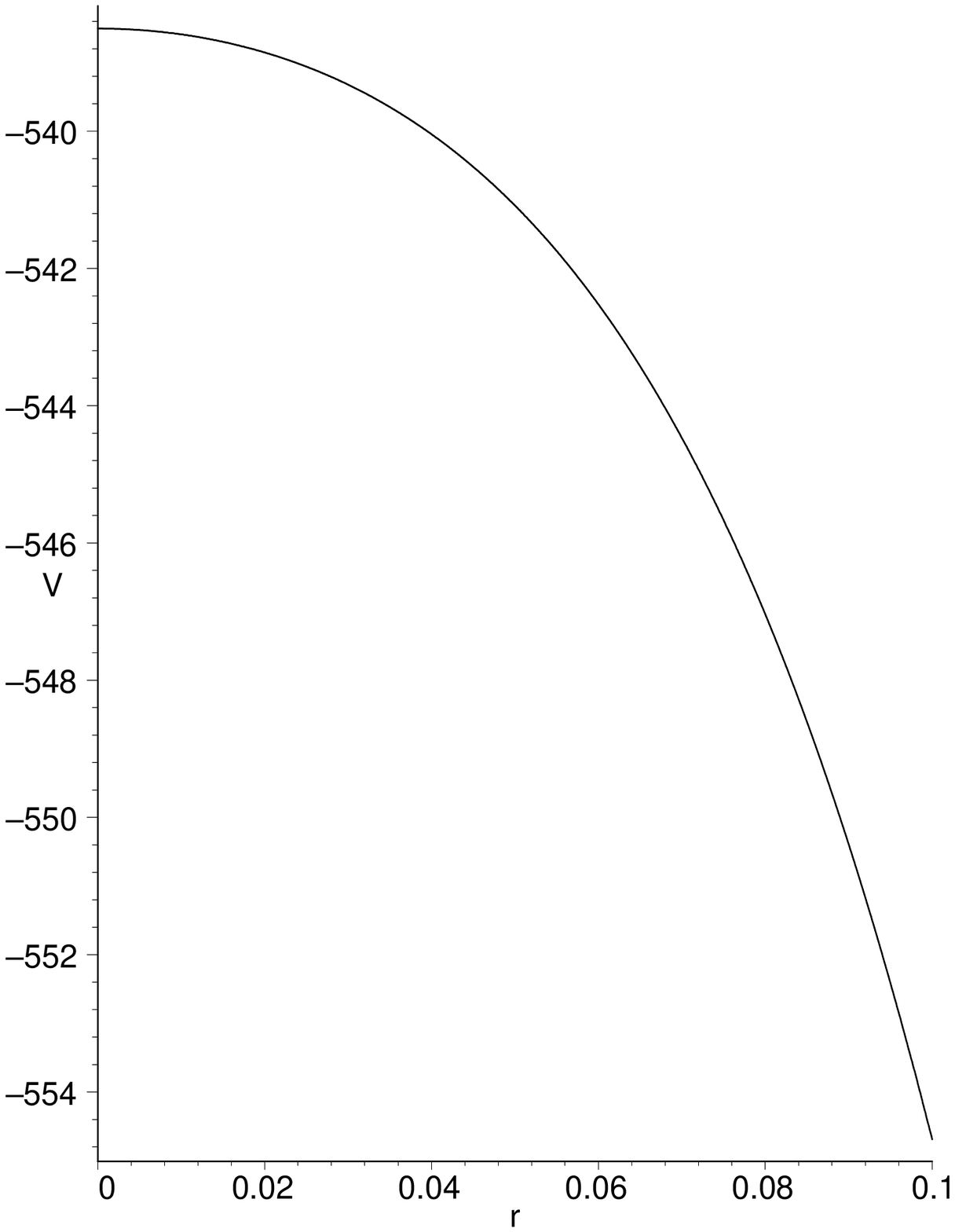} 
\\
\mbox{\footnotesize (a) Potential for $a=0$ at $\kappa = 0.1$} & 
\hspace*{1cm}
\mbox{\footnotesize (b) Potential at $\kappa^2\Lambda^3 = 0.9$}
\end{array} 
\end{eqnarray*} 
\caption{Discrete vacua. 
Parameters are taken to be $(g_0M^0, \Lambda) = (3, 1)$.}
\label{pot-plot}
\end{center}
\end{figure}
In order to see that the potential (\ref{CFpot}) has two vacua 
as local minima, 
we introduce the spherical coordinates to parametrize two 
two-component complex 
fields $\phi_1$ and $\phi_2$ as 
\begin{eqnarray}
\begin{array}{ll}
  \phi_1^1=g(r)\cos(\frac{\theta}{2})\exp(\frac{i}{2}(\Psi+\Phi)), &
\phi_1^2=g(r)\sin(\frac{\theta}{2})\exp(\frac{i}{2}(\Psi-\Phi)), 
\vspace{1mm} \\
\phi_2^1=f(r)\sin(\frac{\theta}{2})\exp(-\frac{i}{2}(\Psi-\Phi)), &
\phi_2^2=-f(r)\cos(\frac{\theta}{2})\exp(-\frac{i}{2}(\Psi+\Phi)) ,
\end{array}
\label{Constsolve}
\end{eqnarray}
where we set 
\begin{eqnarray}
  f(r)^2=\frac{1}{2}(-\Lambda^3+\sqrt{4r^2+\Lambda^6}), \;\;
  g(r)^2=\frac{1}{2}(\Lambda^3+\sqrt{4r^2+\Lambda^6}),
\end{eqnarray}
in order to satisfy the constraint (\ref{CFconst}) 
\cite{IvanovValent}. 
In this coordinate the potential term becomes 
\begin{eqnarray}
 e^{-1}{\cal L}_{pot} 
\!\!&=&\!\!\!\!\!
\frac{-2(g_0M^0)^2}{3(1-\kappa^2\sqrt{4r^2+\Lambda^6})^2} 
\left(v_0+v_1\cos \theta + v_2\cos ^2 \theta \right), \\
  v_0 \!\!&=&\!\!\!\!\!
  3\sqrt{4r^2+\Lambda^6}-\kappa^2(16r^2+3\Lambda^6)
  -4a^2\kappa^2\Lambda^6
  \frac{\sqrt{4r^2+\Lambda^6}-3\kappa^2r^2-\kappa^2\Lambda^6}
    {\sqrt{4r^2+\Lambda^6}-\kappa^2\Lambda^6}, \nonumber \\
    v_1 \!\!&=&\!\!\!\!\! - 8a\kappa^2\Lambda^3
    \frac{r^2+\Lambda^6-\kappa^2\Lambda^6\sqrt{4r^2+\Lambda^6}}
    {\sqrt{4r^2+\Lambda^6}-\kappa^2\Lambda^6}, \;\; 
  v_2 =
    -\Lambda^6\frac{3-2\kappa^2\sqrt{4r^2+\Lambda^6}-\kappa^4\Lambda^6}
    {\sqrt{4r^2+\Lambda^6}-\kappa^2\Lambda^6}. \nonumber
\end{eqnarray}
The scalar potential $V=-e^{-1}{\cal L}_{pot}$ is shown 
in Fig.\ref{pot-plot}. 
There exist two vacua 
at $(r,\theta)=(0,0),(0,\pi)$ 
as local minima 
(see Fig.~1-(a)). 
These two vacua become saddle points with an unstable direction 
along $r$ for $\kappa^2\Lambda^3 > 3/4$ 
 for $a=0$. 
 Fig.~1-(b) shows a typical unstable behavior of potential 
 at $\kappa^2\Lambda^3 =0.9$, 
 which is close to 
$\kappa^2\Lambda^3 =1$, where the target space of hypermultiplets becomes  
a homogeneous space of $SU(2,1)/U(2)$. 
For $a\ne 0$, potential takes different values at these two vacua. 

\vspace{6mm}

{\large\bf BPS equation} 
\vspace{2mm} 
\\
Instead of solving Einstein equations directly, we solve BPS equations  
to obtain a classical solution conserving a half of SUSY. 
Since we consider bosonic configurations, we need to examine 
the on-shell SUSY transformation 
of gravitino and hyperino \cite{Fujita-Ohashi}
\begin{eqnarray}
 \delta_\varepsilon \psi^i_\mu 
  &=& {\cal D}_\mu \varepsilon ^i 
        - \frac{\kappa^2}{6}M_0{\cal Y}_0^{i}{}_j
          \gamma_\mu \varepsilon ^j ,
\label{gravitino1} \\
  \delta_\varepsilon \zeta^\alpha 
   &=& -{\cal D}_\mu {\cal A}^\alpha{}_j\gamma^\mu \varepsilon ^j 
       -( M^1t_1+g_0M^0t_0)^\alpha{}_\beta {\cal A}^\beta{}_j \epsilon^j 
       + \frac{\kappa^2}{2}{\cal A}_j{}^\alpha
        M_0{\cal Y}_0^j{}_k \varepsilon ^k ,
  \label{hyperino1}
\end{eqnarray} 
where 
\begin{eqnarray}
  {\cal D}_\mu \varepsilon^i 
   & =& (\partial_\mu -\frac{1}{4}\gamma_{ab}\omega_\mu^{ab})\varepsilon^i
        - \kappa^2V_\mu{}^i{}_j\varepsilon^j ,  \\
  {\cal D}_\mu {\cal A}_i^\alpha 
   & =& \partial_\mu {\cal A}_i^\alpha +\kappa^2V_\mu{}_i{}^j{\cal A}_j^\alpha 
         -  W^1_\mu t_1^\alpha{}_\beta {\cal A}_i^\beta ,
\label{cov-deriv} \\
  {\cal Y}_0^{ij} &=& 2{\cal A}_\alpha^{(i}d^\alpha{}_\gamma
                (g_0t_0)^{\gamma \beta}{\cal A}^{j)}_\beta, \quad 
  V_\mu^{ij} = 
   - {\cal A}^{\gamma (i}d_\gamma^{~\alpha}\nabla_\mu{\cal A}^{j)}_\alpha .
\end{eqnarray}
If we assume the warped metric (\ref{5dmetric}),  
the SUSY transformation of the gravitino (\ref{gravitino1}) 
decouples into two parts 
\begin{eqnarray} 
  \delta_\varepsilon \psi_m^i 
   &=& \partial_m \varepsilon^i 
       - \frac{1}{2}\gamma_m \gamma^y \partial_y U\cdot \varepsilon^i 
       - \frac{\kappa^2}{6}M^0{\cal Y}_0^i{}_j
         \gamma_m \varepsilon^j ,
  \label{gravitino2a} \\
  \delta_\varepsilon \psi_y^i 
   &=& \partial_y \varepsilon^i - \kappa^2 V_y^i{}_j \varepsilon^j
        - \frac{\kappa^2}{6}M^0{\cal Y}_0^i{}_j
          \gamma_y \varepsilon^j .
  \label{gravitino2b}
\end{eqnarray}

Let us require vanishing of the SUSY variation of gravitino and hyperino 
to preserve four SUSY specified by 
\begin{eqnarray}
  \gamma^y \varepsilon^i(y) = i \tau_3^{i}{}_j \varepsilon^j(y),
\label{1/2SUSY}
\end{eqnarray}
where $\tau_3$ is one of the Pauli matrix.
Then one of the gravitino BPS conditions (\ref{gravitino2a}) 
gives an equation for the warp factor $U(y)$ and an 
additional constraint 
\begin{eqnarray}
  \partial_y U = {\cal W}(\phi)&\!\!\!\equiv &\!\!\! 
        \frac{2\kappa^2}{3}g_0M^0\bar{\cal A}^{-1}
       [-a\Lambda^3-(\phi_1^*\sigma_3\phi_1-\phi_2^*\sigma_3\phi_2)] ,
\label{BPS-warp}\\
  \phi_1^* \sigma_3 \phi_2 &\!\!\!=&\!\!\! 0 .
\label{add-const}
\end{eqnarray}
The hyperino BPS condition (\ref{hyperino1}) 
combined with the condition (\ref{add-const}) gives 
\begin{eqnarray}
 \left[
   {\partial}_y -i W^1_y+\left(\frac{3}{2}{\cal W}(\phi)+\bar{V}\right)
    + (
-g_0M^0
\sigma_3+ M^1)
 \right](\bar{\cal A}^{-\frac{1}{2}}\phi_1) &=& 0 , \nonumber \\
 \left[
   {\partial}_y -i W^1_y+\left(\frac{3}{2}{\cal W}(\phi)-\bar{V}\right)
    - (
-g_0M^0
\sigma_3+ M^1)
 \right](\bar{\cal A}^{-\frac{1}{2}}\phi_2) &=& 0 ,
\label{BPS-hyper}
\end{eqnarray}
\begin{equation}
\bar{V}
  \equiv  \kappa^2\bar{\cal A}^{-1} 
          (\phi_1^*\overset{\leftrightarrow}{\partial}_2\phi_1
          -\phi_2^*\overset{\leftrightarrow}{\partial}_2\phi_2)/2 .
\label{V-onshell}
\end{equation}
Since Eq.(\ref{1/2SUSY}) assures that solutions of these BPS 
equations conserve four SUSY out of eight SUSY, the effective theory on 
this background has ${\cal N}=1$ SUSY in four dimensions. 
This should be useful for 
model building in the 
SUSY brane-world scenario. \\

\vspace{4mm}

{\large\bf Wall solution 
and thin wall limit} 
\vspace{2mm}
\\
Let us rewrite 
the BPS equations 
in terms of the spherical coordinates (\ref{Constsolve}). 
After some algebra, we obtain four independent 
differential equations from Eqs.~(\ref{BPS-hyper}), 
\begin{eqnarray}
r \frac{d \Psi}{d y} &=& 0, \quad 
\frac{d r}{d y} = \frac{2g_0M^0\sqrt{4r^2+\Lambda^6}}
                  {\sqrt{4r^2+\Lambda^6}-\kappa^2\Lambda^6}
                  \cdot r(\cos \theta +a\kappa^2\Lambda^3), \nonumber \\
\sin {\theta }\frac{d \Phi}{d y} &=& 0, \quad 
\frac{d \theta}{d y} = -2g_0M^0\sin \theta .
\label{BPS-spherical}
\end{eqnarray}
Let us obtain the wall solution interpolating between the two vacua: 
 $(r, \theta)=(0,0),(0,\pi)$. The boundary condition of $r=0$ 
at $y=-\infty$ dictates the solution of (\ref{BPS-spherical}) as 
\begin{eqnarray}
 r=0, \;\; \cos \theta = \tanh \left(2g_0M^0(y-y_0)\right),\;\; 
\Phi=\varphi_0,
\label{BPSsol-spherical}
\end{eqnarray}
 with $\Psi$ undetermined, and $y_0$ and $\varphi_0$ are constants.
Substituting these solutions to r.h.s.~of Eq.(\ref{BPS-warp}), 
we obtain the BPS solution of the warp factor  
\begin{eqnarray}
 U(y) = -\frac{\kappa^2\Lambda^3}{3(1-\kappa^2\Lambda^3)}
        \left[ \ln \{\cosh \left(2g_0M^0(y-y_0)\right)\} 
           + 2ag_0M^0(y-y_0)\right].
\label{Warp-sol}
\end{eqnarray}
The warp factor $e^{2U(y)}$ of this solution 
decreases exponentially for both infinities $y\rightarrow \pm \infty$ 
for $|a|<1$ (see Fig.~2) similarly to the case of the bulk AdS space. 
Therefore a four-dimensional massless graviton should be 
localized on the wall \cite{CsabaCsaki}. 
The cases of 
$|a|=1$ become the wall solutions interpolating 
between AdS and flat Minkowski vacua. 
On the other hand, warp factor increases exponentially either one of the 
infinities for $|a|>1$. 
 {}Following the AdS/CFT conjecture, a vacuum reached by a decreasing 
(increasing) warp factor corresponds to IR (UV) fixed point of a 
four-dimensional field theory \cite{AdS/CFT}. 
Our BPS wall solutions interpolate two IR fixed points for 
$|a|<1$. 
Moreover these vacua are local minima of the potential. 
This implies that no relevant operator exists in these 
conformal field theories\footnote{
One of the authors (NS) thanks Steve Gubser for a discussion on this point. 
}. 
The wall solutions for $|a|>1$ interpolate one IR and one UV fixed points 
which cannot realize the warped 
extra dimension, but should be related to a Renormalization 
Group (RG) flow : 
the function ${\cal W}(\phi)$ in BPS equation of warp factor (\ref{BPS-warp}) 
is monotonic without changing its sign along the flow. 
The family of our BPS solutions contains a parameter $a$ interpolating 
between three classes of field theories : 
one with two IR fixed points ($|a|<1$), 
another with one IR and one UV fixed point ($|a|>1$), 
and one with one IR fixed point and flat space ($|a|=1$). 
We find it remarkable that a 
single family of models can realize all these possibilities 
as we change a parameter. 

\begin{figure}[t]
\begin{center}
\leavevmode
\begin{eqnarray*}
\begin{array}{ccc}
  \epsfxsize=4.5cm
  \epsfysize=3.5cm
\epsfbox{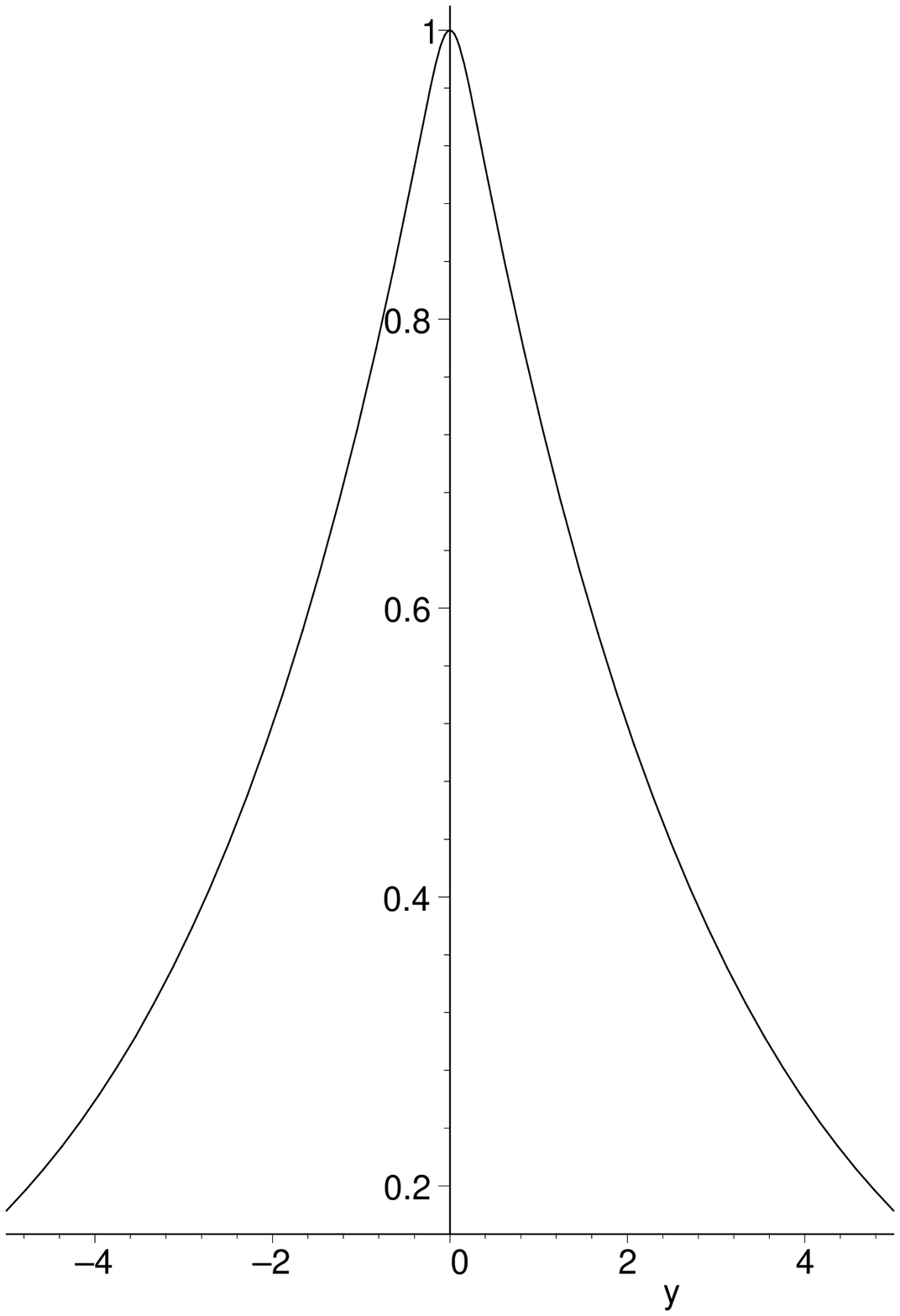} & 
  \epsfxsize=4.5cm
  \epsfysize=3.5cm
\hspace*{1cm}
\epsfbox{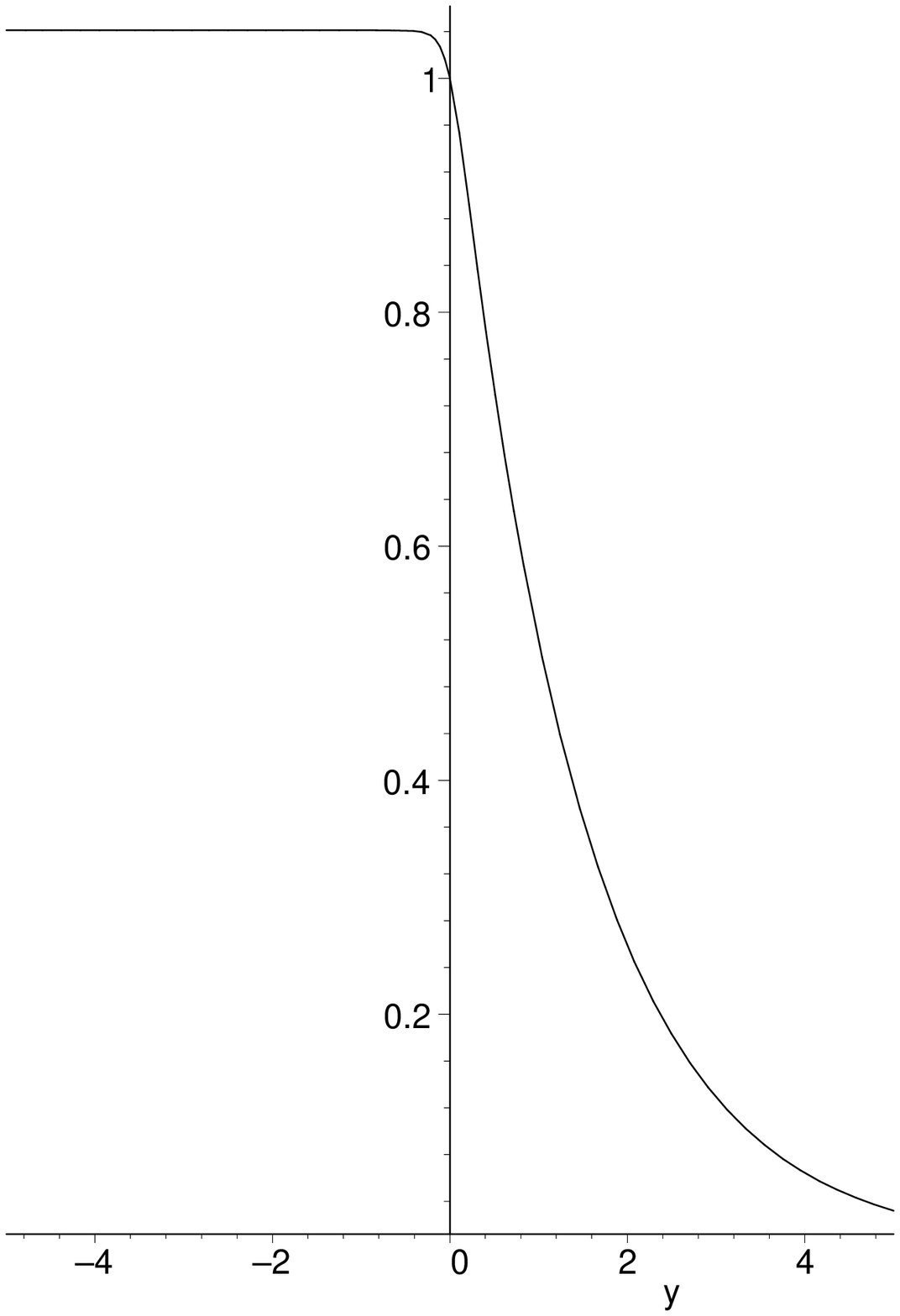} %
  \epsfxsize=4.5cm
  \epsfysize=3.5cm
\hspace*{1cm}
\epsfbox{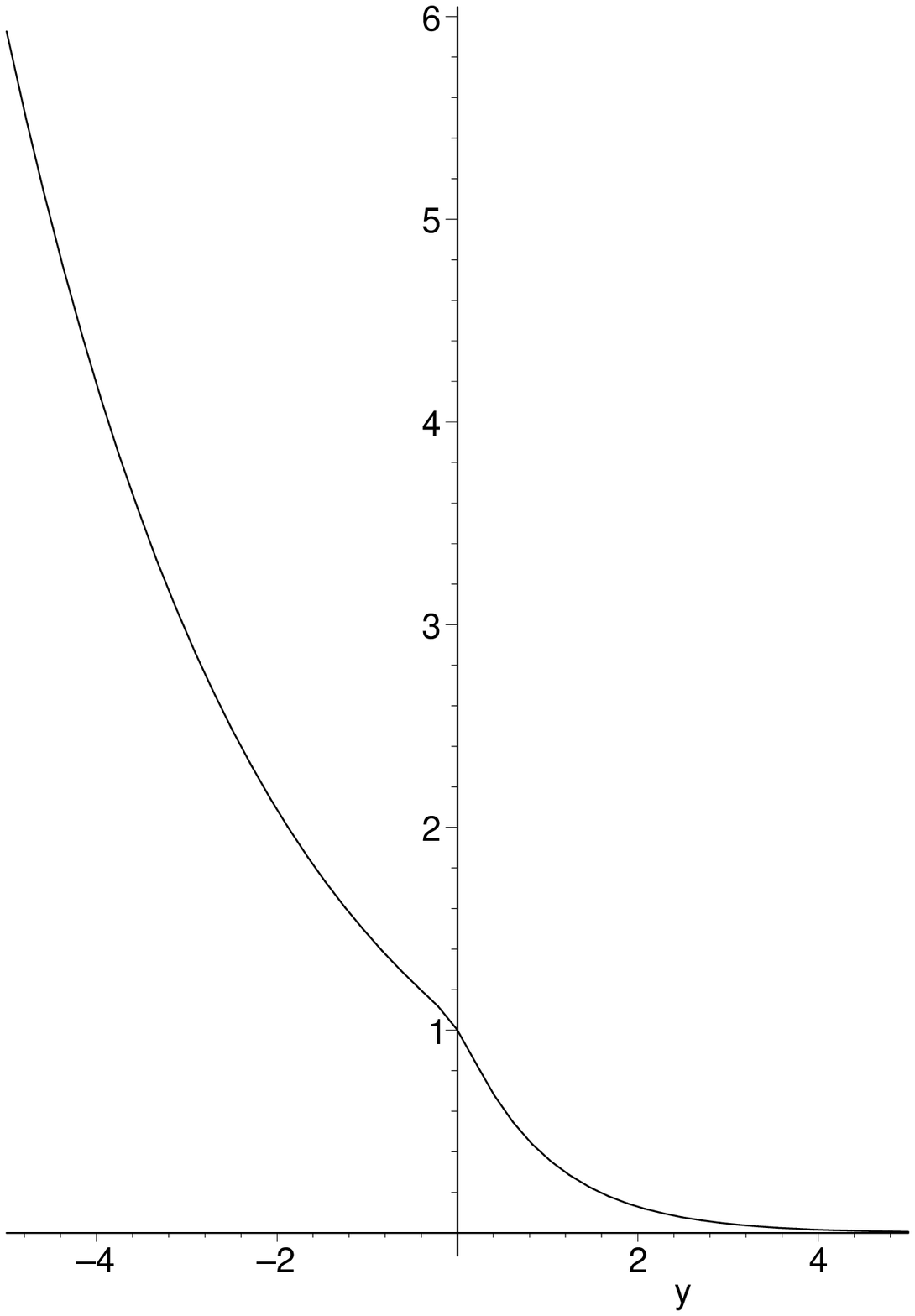} 
\hspace{3cm}
\\
\hspace{0cm}\mbox{\footnotesize (a) a=0} &
\hspace{-8cm}\mbox{\footnotesize (b) a=1}  &
\hspace{-6cm}\mbox{\footnotesize (c) a=2} 
\end{array} 
\end{eqnarray*} 
\caption{Profile of warped metric. 
Parameters are taken to be $(g_0M^0, \Lambda, \kappa) = (3, 2, 0.1)$}
\end{center}
\end{figure}
 From Eqs.(\ref{gravitino2b}) and (\ref{1/2SUSY}) 
we find the Killing spinor $\varepsilon^i(y)$ as
\begin{eqnarray}
  \varepsilon^i(y)\equiv e^{U(y)/2}\tilde{\varepsilon}^i, \quad 
\gamma^y \tilde{\varepsilon}^i = i\tau_{3}^{i}{}_j\tilde{\varepsilon}^j,
\label{killing-sol}
\end{eqnarray}
where $\tilde{\varepsilon}^i$ is a constant spinor.

We can obtain a thin wall limit by taking $g_0M^0 \rightarrow \infty$ 
and $\Lambda \rightarrow 0$ with $g_0M^0\Lambda^3$, $\kappa$, 
and $a$ fixed. 
Substituting the solutions (\ref{BPSsol-spherical}) and (\ref{Warp-sol}) 
to the Lagrangian of hypermultiplets and taking the thin wall limit, 
we obtain for $y_0=0$ 
\begin{eqnarray}
  &\!\!\!&\!\!\!-\frac{1}{2\kappa^2}R + e^{-1}{\cal L}_{kin} 
    + e^{-1}{\cal L}_{pot} \nonumber \\
&\!\!\!=&\!\!\! -\frac{1}{2\kappa^2}R 
+\frac{8\kappa^2(g_0M^0\Lambda^3)^2}{3(1-\kappa^2 \Lambda^3)^2}
 \left[a+\tanh (2g_0M^0y) \right]^2 \!\!
  - {2g_0M^0\Lambda^3(2-\kappa^2\Lambda^3) \over (1-\kappa^2 \Lambda^3)^2}
  {g_0M^0 \over (\cosh(2g_0M^0y))^2} \nonumber \\
&\!\!\!{}&\!\!\! \to -\frac{1}{2\kappa^2}R 
  - \left[ -\frac{8\kappa^2(g_0M^0\Lambda^3)^2}{3}\left(a+\epsilon(y)\right)^2
         \right]
  - 4(g_0M^0\Lambda^3)
    \cdot \delta (y) ,
\end{eqnarray}
where $\epsilon(y)\equiv \pm 1$ is a sign function. 
We have obtained a boundary cosmological constant from the wall tension 
$T_w$ and a bulk cosmological constant $\Lambda_c^{+}, (\Lambda_c^{-})$ 
for $y<0 (y>0)$ 
as 
\begin{eqnarray}
  T_w = 4(g_0M^0\Lambda^3),
 \qquad 
   \Lambda_c^{\pm} 
      = -\frac{8\kappa^2(g_0M^0\Lambda^3)^2}{3}(1 \pm a)^2 .
\end{eqnarray}
We find that  our BPS solution for $a=0$ 
automatically satisfies 
 the fine-tuning condition $\sqrt{-\Lambda_c}=\frac{\kappa}{\sqrt{6}}T_w$ 
 of the Randall-Sundrum model between 
$T_w$ and 
$\Lambda_c$, 
as a result of combined dynamics of scalar field and 
gravity. 
In terms of the asymptotic linear exponent $c$ of the warp factor 
$ U\sim - c |y-y_0|, \; c \equiv 2\kappa^2(g_0M^0\Lambda^3)/3$ for 
$|y-y_0|\to \infty$,  
the wall tension 
$T_w =24c/(4\kappa^2)$,  and cosmological constant 
 $\Lambda_c = -24 c^2/(4\kappa^2)$ 
satisfy precisely the same 
relation as in Ref.\cite{RS2} 
(with $M_p^3\equiv (4\kappa^2)^{-1}$). 
Therefore we have realized the single-wall Randall-Sundrum model 
as a thin-wall limit of our solution of the coupled scalar-gravity theory, 
instead of an artificial boundary cosmological constant put 
at an orbifold point. 

By a dimensional reduction, we can obtain 
from the above hypermultiplet action 
an ${\cal N}=2$ four-dimensional supergravity theory (eight SUSY) 
with hypermultiplets. 
Therefore we can automatically obtain 
from our BPS wall solution (\ref{BPSsol-spherical})-(\ref{Warp-sol}) 
a BPS 
wall solution 
in ${\cal N}=2$ four-dimensional supergravity 
which is a gravitational deformation of the BPS wall solution 
 \cite{Townsend}, \cite{ANNS} in 
the global SUSY case.

\vspace{6mm}

{\large\bf Weak gravity limit} 
\vspace{2mm}
\\
%
%
Next, we discuss the properties of our model and solution 
in the weak gravity limit, 
which is defined by taking 
the limit of $\kappa \to 0$ with $g_0M^0\equiv \bar{M}$ 
held fixed. We obtain in the limit  
\begin{eqnarray}
\frac{1}{2}e^{-1}({\cal L}_{kin}+{\cal L}_{pot})
  &\to & -(\partial^\mu \phi_1^*\partial_\mu \phi_1
            + \partial^\mu \phi_2^*\partial_\mu \phi_2) 
            +(|\phi_1|^2+|\phi_2|^2)W_\mu^1 W^{1\mu} \nonumber \\
   &{}&   -\bar{M}^2\frac{(|\phi_1|^2+|\phi_2|^2)^2
         - (\phi_1\sigma_3\phi_1^*+\phi_2^*\sigma_3\phi_2)^2}
           {|\phi_1|^2+|\phi_2|^2} ,
\label{globallimit} \\
 W_\mu^1 &\to & 
\frac{i(\phi_1\overset{\leftrightarrow}{\partial}_\mu \phi_1^* 
+ \phi_2\overset{\leftrightarrow}{\partial}_\mu \phi_2^*)}
{2(|\phi_1|^2+|\phi_2|^2)} ,
\end{eqnarray}
and the constraints (\ref{CFconst}) are unchanged.
The kinetic part in Eq.(\ref{globallimit}) 
is identical to the five-dimensional version of the 
nonlinear sigma model with 
the target space of $T^*{\bf C}P^1$, namely the Eguchi-Hanson manifold, 
in the basis of 
Curtright and Freedman \cite{CF}.   
The potential term is also identical to the mass term for 
this nonlinear sigma model, as discussed in Ref.\cite{ANNS}:
this mass term is originated from the central extension of 
global ${\cal N}=2$ SUSY algebra \cite{ANNS}, and can be rewritten as 
the norm of a tri-holomorphic killing vector for an isometry of 
target space of the Eguchi-Hanson metric \cite{Townsend}. 
Therefore the above action (\ref{globallimit}) 
has global ${\cal N}=2$ SUSY. 

In this limit, BPS equations for scalar fields in the hypermultiplets become 
\begin{eqnarray}
r\frac{d \Psi}{d y} = 0, \;\;
\frac{d r}{d y} = 2\bar{M}r\cos \theta, \;\;
\sin {\theta} \frac{d \Phi}{d y} = 0, \;\; 
\frac{d \theta}{d y} = -2\bar{M}\sin \theta .
\end{eqnarray}
These equations are identical to the BPS equations in the massive 
Eguchi-Hanson sigma model, whose four-dimensional version has been 
discussed in Ref.\cite{ANNS}. 
Therefore the model and the solution we discuss in this paper 
are consistent gravitational deformation 
of the massive Eguchi-Hanson nonlinear sigma model in five dimensions 
and associated BPS wall solutions. 

The wall solution for $\kappa=0$ is the five-dimensional version of 
the kink solution in Ref.\cite{Townsend} with the field redefinition 
in Ref.\cite{ANNS}. 
Their solutions 
 are exactly identical to our solution 
(\ref{BPSsol-spherical}) obtained for finite $\kappa$. 
It is very interesting that 
BPS solution for the hypermultiplet $\phi$ 
in the global SUSY model coincides with 
that in the corresponding supergravity. 
This mysterious coincidence has also appeared in the analytic 
solution in a four-dimensional ${\cal N}=1$ supergravity 
model \cite{EMSS}.  
It is tempting to speculate that this property 
might be related to the exact solvability of our model. 

It has been a long-standing problem to find a consistent gravitational 
deformation from a hyper-K\"ahler manifold to a quaternionic K\"ahler manifold 
with gravitationally corrected potential terms necessary for wall solutions. 
We have achieved this goal by using an off-shell formulation of supergravity and 
the {\it massive} quaternionic K\"ahler quotient method\footnote{
Massless quaternionic K\"ahler quotient method has been used before 
 \cite{Galicki}, \cite{IvanovValent}. 
}. 
Supergravity domain walls have been extensively worked out 
using the on-shell formulation such as 
in Ref.\cite{Cere-Dall}.
Since auxiliary fields are eliminated when we solve 
BPS equations, 
it should in principle be possible to obtain BPS solutions 
from the on-shell formulation. 
Off-shell formulation of supergravity, however, offers a more 
powerful tool to obtain supergravity domain walls 
as gravitational deformations of those in global SUSY models. 
If we eliminate constraints before coupling to gravity, 
it is very difficult in general to extend hyper-K\"ahler nonlinear 
sigma models 
with global eight SUSY to quaternionic K\"ahler nonlinear 
sigma models coupled to supergravity, 
because of the complicated gravitational corrections. 
On the other hand, many hyper-K\"ahler sigma models can be obtained 
as quotients of linear sigma models by using vector multiplets as 
Lagrange multipliers. 
When we eliminate Lagrange multiplier multiplets after coupling to 
gravity in the off-shell formulation, 
we obtain quaternionic K\"ahler nonlinear sigma models 
coupled to supergravity. 
Moreover we can take a weak gravity limit of these models straightforwardly.
Therefore the off-shell formulation of supergravity is quite useful 
to obtain quaternionic nonlinear sigma models as continuous 
gravitational deformations of hyper-K\"ahler nonlinear sigma models 
of the global SUSY. 

As noted in Ref.\cite{IvanovValent}, our quaternionic manifold has a 
conical singularity at $r=0$ in $r, \Psi$ plane except for discrete values 
of gravitational coupling $\kappa^2\Lambda^3=(k-1)/k, \; k=2,3,\dots$ 
where it can be identified with a removable bolt singularity. 
Our BPS solution can be realized for 
this smooth manifold at least 
for $k=2,3$ ($k=2$) for $a=0$ ($|a|>1$) 
without having saddle points of the scalar potential. 
Moreover we believe that we can 
achieve a continuous 
gravitational deformation avoiding the singularity, 
if we simply restore 
the finite gauge coupling $g_1$ for the vector multiplet containing $W_\mu^1$ 
instead of infinite gauge coupling as we did up to now. 
Let us take a gauged linear sigma model consisting of 
hypermultiplets interacting with vector multiplets and 
couple it to supergravity by the tensor calculus \cite{Fujita-Ohashi}. 
This model is a perfectly consistent interacting supergravity 
system with eight local SUSY. 
 For finite but large values of gauge coupling $g_1$, it effectively reduces 
 to our quaternionic nonlinear sigma model except near the 
 conical singularity where we can no longer neglect the vector multiplet. 
 Only in the neighborhood of the singularity, the manifold loses its simple 
 geometrical meaning of quaternionic manifold consisting solely of 
 hypermultiplets. 
We may call this situation a resolution of the conical 
singularity\footnote{
One of the authors (NS) thanks Tohru Eguchi for an illuminating 
discussion on this point. 
} 
in the spirit of Ref.\cite{Witten}. 
In this 
model, we can freely take 
the limit $\kappa \rightarrow 0$ to obtain the Eguchi-Hanson manifold. 
Therefore we believe that this gauged linear sigma model coupled 
with supergravity 
is the most appropriate setting for the gravitational deformation of 
hyper-K\"ahler manifolds such as Eguchi-Hanson manifold. 
On the other hand, our BPS wall should still be a valid solution of the 
gauged linear sigma model  coupled with supergravity. 
This is because 
our constraints arising from the 
elimination of the vector multiplet without kinetic 
term 
preserve all SUSY, and hence they 
solve the BPS condition 
for the vector multiplet 
trivially. 
Therefore we anticipate that our solution continues to be a BPS wall solution 
for the gauged linear sigma model with a finite large coupling $g_1$ 
coupled with supergravity. 
The only modification should be that 
the vector multiplet cannot be neglected 
when we examine the geometry of the target manifold 
near the resolved conical singularity. 
We hope to provide a full analysis of the gauged linear sigma model 
coupled with supergravity 
in subsequent publications. 

\vspace{4mm}

{\large\bf Discussion 
} 
\vspace{2mm} 
\\
 Finally we discuss implications of our solution 
 on two no-go theorems. It has been shown 
that wall solutions in supergravity theories always have singularities 
under several assumptions including non-positive scalar potential 
\cite{MaldacenaNunez}. 
Our BPS wall solution has no singularities of the type they discussed 
and can be regarded as a counter example of the no-go theorem. 
This violation of the no-go theorem arises from the fact that 
a potential becomes positive around the center of the wall 
contradicting one of their assumptions. 

On the other hand, it has been shown that 
the proposed Nambu-Goldstone (NG) fermion from broken SUSY 
diverges on the wall in supergravity theories \cite{Gibbons-Lam}. 
They considered the result as a no-go theorem for smooth 
wall solutions such as the one presented here. 
Recently Cvetic and Lambert have proposed a more proper definition of 
the wave function of the NG fermion associated with the 
killing spinor of broken SUSY 
and have argued that the no-go theorem can be evaded \cite{Cvetic-Lam}. 
In fact, 
we obtained an explicit domain wall solution with warp factor decreasing 
for both infinities of extra dimensions in five-dimensional supergravity. 
We regard our result to be an example evading the no-go theorem 
along the line of Ref.\cite{Cvetic-Lam}.

\vspace{2mm}

\begin{center}
{\bf Acknowledgements}
\end{center}
We are grateful for discussion with Keisuke Ohashi. 
 One of the authors (MN) thanks useful discussion with Koji Hashimoto 
and another (NS) thanks useful discussion with Tohru Eguchi, Steve Gubser, 
Sergei Ketov and Taichiro Kugo. 
This work is supported in part by Grant-in-Aid 
 for Scientific Research from the Japan Ministry 
 of Education, Science and Culture 13640269. 
The work of M.~Naganuma is supported by JSPS Fellowship. 


\end{document}